# Photon-Counting Spectral Phase-Contrast Mammography

E. Fredenberg,[a,b] E. Roessl,[c] T. Koehler,[c] U. van Stevendaal,[c] I. Schulze-Wenck,[d]
N. Wieberneit,[d] M. Stampanoni,[e,f] Z. Wang,[e] R. A. Kubik-Huch,[g] N. Hauser,[h] M. Lundqvist,[b]
M. Danielsson,[a,b] M. Åslund[b]

[a]Department of Physics, Royal Institute of Technology (KTH), AlbaNova University Center,
106 91 Stockholm, Sweden;

[b]Philips Women's Healthcare, Smidesvägen 5, 171 41 Solna, Sweden;

[c]Philips Technologie GmbH Innovative Technologies, Research Laboratories, Röntgenstrasse
24, 22335 Hamburg, Germany;

[d]Philips Healthcare, Röntgenstrasse 24, 22335 Hamburg, Germany;

[e]Swiss Light Source, Paul Scherrer Institut, 5232 Villigen, Switzerland;

[f]Institute for Biomedical Engineering, University and ETH Zürich, 8092 Zürich, Switzerland;

[g]Department of Radiology, Kantonsspital Baden, 5404 Baden, Switzerland;

[h]Department of Gynecology and Obstetrics, Interdisciplinary Breast Center Baden,
Kantonsspital Baden, 5404 Baden, Switzerland

## ABSTRACT

Phase-contrast imaging is an emerging technology that may increase the signal-difference-to-noise ratio in medical imaging. One of the most promising phase-contrast techniques is Talbot interferometry, which, combined with energy-sensitive photon-counting detectors, enables spectral phase-contrast imaging. We have evaluated a realistic system for spectral phase-contrast mammography by cascaded-systems analysis. Phase contrast improved detectability compared to absorption contrast, in particular for fine tumor structures. This result was supported by images of human mastectomy samples that were acquired with a conventional detector. The optimal incident energy was slightly higher in phase contrast than in absorption contrast. In Talbot interferometry, the optimum was sharp and located at the setup design energy. Detectability may be further improved by energy weighting, which for phase contrast was found to have a weaker energy dependence than absorption contrast. Optimal weighting in Talbot interferometry had a steeper energy dependence, with a maximum at the setup design energy. Spectral material decomposition was not facilitated by phase contrast, but phase information may be used instead of spectral information.

**Keywords:** mammography; phase contrast; spectral imaging; detectability index; Talbot interferometry; photon counting

## 1. INTRODUCTION

Phase-contrast imaging is an emerging technology in medical x-ray imaging that may increase the signal-difference-to-noise ratio compared to conventional absorption contrast.[1–3] One of the most promising phase-contrast techniques for medical imaging is Talbot interferometry.[4–8] Benefits of Talbot interferometry in a medical imaging context include low coherence requirements, a compact setup, phase and absorption contrast are readily separated, and good photon economy. A challenge of the technique is that Talbot interferometers are optimized only at a single energy – the design energy.

---

In absorption contrast, there exists an optimal incident energy because of the tradeoff between contrast and dose, which both increase towards lower energies.[9] The same kind of tradeoff exists in phase-contrast imaging with a resulting optimal energy.[10]

Photon-counting detectors are fast enough to measure the energy of individual photons[11] and can be employed for single-shot spectral imaging. There are two well-investigated applications of spectral imaging: (1) Energy weighting is optimization of the signal-to-noise ratio by weighting photons according to information content.[12–15] (2) Material decomposition extracts information about the object constituents by the material-specific absorption energy dependence.[16,17] A special case is energy subtraction, which aims at reducing the impact of anatomical-structure overlap, so-called anatomical noise.[14,15,18–21]

Spectral absorption-contrast imaging is well investigated, whereas spectral phase contrast is a relatively unexplored area. We evaluate two aspects of the x-ray energy spectrum on phase contrast in general, and on a mammography system based on photon-counting Talbot interferometry in particular: (1) optimization of the incident spectrum with respect to energy, and (2) utilization of the transmitted spectrum (spectral imaging). The framework that we will use for evaluation has been presented previously,[22] and is introduced for non-spectral imaging.

## 2. MATERIAL AND METHODS

### 2.1. A photon-counting phase-contrast mammography setup

We have investigated a photon-counting Talbot-interferometry setup for phase-contrast mammography with silicon strip detectors and geometry similar, but not identical, to the Philips MicroDose Mammography system (Philips Digital Mammography AB, Solna, Sweden).[23,24] The setup will be further described in an upcoming publication,[22] but the main components are outlined in Fig. 1 and described in the following under the assumption of a parallel beam.

A $\pi$-shifting beam splitter (phase grating) introduces interference fringes at $D_n = np_1^2/8\lambda^d$, where $n = 1, 3, 5\ldots$ is the Talbot order, $p_1$ is the beam-splitter pitch, and $\lambda^d$ is the wave length for which the setup is designed (the design wave length).[7] A phase gradient in the object causes a phase shift (displacement) of the fringes, which can be measured in the direction perpendicular to the grating slits ($x$) to obtain the differential phase shift that is caused by the object. Absorption contrast can be measured in parallel to phase contrast by averaging over the fringes.

The fringe period is generally very short (in the order of $p_1/2$), but a fine-pitch analyzer grating can be scanned over the fringe pattern to demodulate the high-frequency fringes into lower frequencies, which can be detected by much larger detector pixels (phase stepping). To improve photon economy, a relatively large source may be covered by a grating with a pitch such that Talbot images from different grating slits coincide (Talbot-Lau geometry).[7]

### 2.2. Non-spectral imaging

#### 2.2.1. Cascaded-systems analysis

We have developed a cascaded-systems framework to evaluate the performance of phase-contrast imaging that will be presented in an upcoming publication.[22] The framework is based on the noise-equivalent number of quanta (NEQ) and an ideal-observer detectability index ($d'$):[25–29]

$$\mathrm{NEQ}(\boldsymbol{f}) = \frac{\langle I\rangle^2 T^2(\boldsymbol{f})}{S(\boldsymbol{f})} \quad \text{and} \quad d'^2 = \int_{\mathrm{Ny}} \mathrm{NEQ}(\boldsymbol{f}) \times W^2(\boldsymbol{f})\,\mathrm{d}\boldsymbol{f}, \tag{1}$$

where $\boldsymbol{f}$ is the spatial frequency vector, $\langle I\rangle$ is the expected image signal, $T$ is the modulation transfer function (MTF) of the system, and $S$ is the noise-power spectrum (NPS). $W = C \times F$ is the task function with contrast $C$ and signal template $F$.

To compare phase and absorption contrast side-by-side, the differential phase-contrast signal was integrated in the $x$ direction, and the absorption-contrast signal was taken as the logarithm of the detected number of photons. Hence, the signal difference between target material $c$ and background material $g$ in phase contrast (subscript $\Phi$) and absorption contrast (subscript $\mu$) is[22]

$$\Delta s_\Phi = |\langle I_{\Phi c}\rangle - \langle I_{\Phi g}\rangle| = k|\delta_c - \delta_g|d_c \equiv k\Delta\delta_{cg}d_c \quad \text{and} \quad \Delta s_\mu = |\langle I_{\mu c}\rangle - \langle I_{\mu g}\rangle| = |\mu_c - \mu_g|d_c \equiv \Delta\mu_{cg}d_c, \quad (2)$$

where $k = 2\pi/\lambda$ is the wave number; $\delta_c$ and $\delta_g$ are the real parts of the complex refractive index for the respective materials; $\mu$ is the linear attenuation coefficient; $\Delta\delta_{cg}$ and $\Delta\mu_{cg}$ are the differences in $\mu$ and $\delta$; $d_c$ is the target thickness.

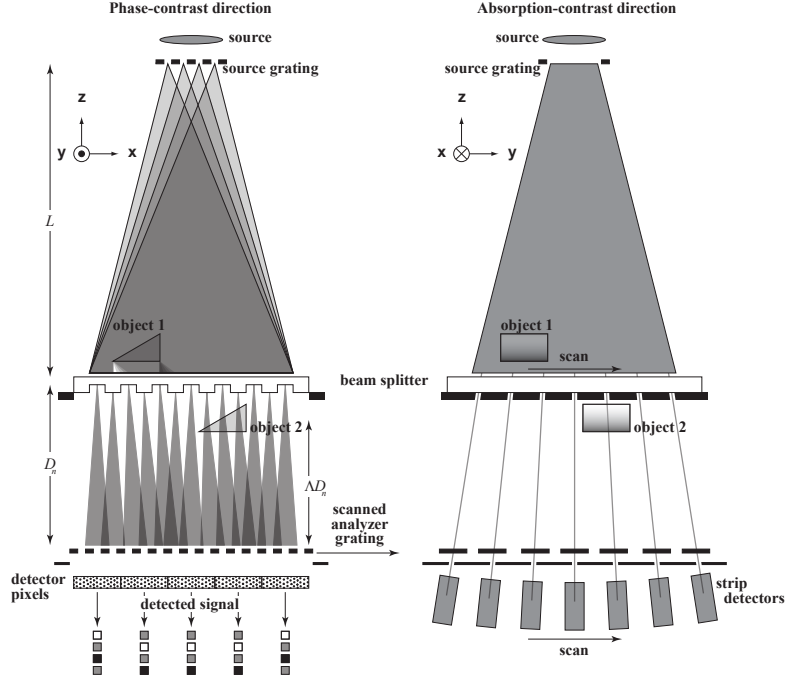

**Figure 1.** Schematic of the photon-counting Talbot-interferometry setup for phase-contrast mammography. **Left:** A beam splitter (phase grating) illuminated by an x-ray source that is covered by a source grating induces interference fringes. The fringes are displaced by the phase gradient in an object. A fine-pitch analyzer grating can be used to demodulate the high-frequency fringes into lower frequencies so that the fringe displacement and hence the phase shift can be measured by the photon-counting silicon strip detector elements. **Right:** The silicon strip detectors are scanned in the other direction to cover the full field-of-view.

The MTF in phase and absorption contrast was derived as[22]

$$T_\Phi(\boldsymbol{f}) = T_0(\boldsymbol{f}) \quad \text{and} \quad T_\mu(\boldsymbol{f}) = T_0(\boldsymbol{f}), \quad (3)$$

where subscript 0 indicates the system MTF. The MTF is hence equal for phase and absorption contrast. The NPS on the other hand differs, and because of the one-dimensional detection of the phase derivative, the phase-contrast NPS differs in the $x$- and $y$-directions:[22]

$$S_{Q\Phi}(\boldsymbol{f}) = \begin{cases} \dfrac{8}{\pi^2} \times \dfrac{1}{\Lambda^2 n^2 p_1^2} \times \left[\dfrac{\lambda^d}{\lambda}\right]^2 \times \dfrac{1}{A^2} \times \dfrac{1}{f_x^2} \times \dfrac{1}{N^2} \times S_{Q0}(f_x) \\ \dfrac{2}{N^2} \times S_{Q0}(f_y) \end{cases} \quad \text{and} \quad S_{Q\mu}(\boldsymbol{f}) = \dfrac{1}{N^2} \times S_{Q0}(\boldsymbol{f}), \quad (4)$$

where $N$ is the number of photons incident on the analyzer grating. $\Lambda \in (0,1)$ ranges from 0 for an object at the detector to 1 for an object at or upstream of the beam splitter. Hence, the object may be located after the beam splitter ($\Lambda < 1$) to keep the setup compact and avoid beam-splitter absorption, but the cost is increased noise (equivalent to reduced phase-contrast sensitivity). Moreover, problems arise with thick objects that are placed after the beam splitter since the displacement of the interference fringes will depend on distance from the detector. $A \in (0,1)$ is the amplitude of the interference fringes (often referred to as "visibility"), which is affected by coherence. The inverse frequency dependence of $S_{Q\Phi}$ in the $x$-direction is caused by the integration from differential phase contrast to phase contrast. The inverse wavelength dependence is induced by the smaller angular deviation at shorter wavelengths.

Moreover, a phase-propagation simulation framework was implemented in order to validate the analytical calculations.[22]

### 2.2.2. Phase-contrast imaging of human mastectomy breast samples

In a recently published study native, human breast mastectomy samples were scanned with a prototype Talbot interferometer designed at the Paul-Scherrer Institute (PSI) in Villigen, Switzerland.[30] After radical mastectomy at the Interdisciplinary Breast Center, Baden, Switzerland, the samples were transported in a dedicated cooled sample holder to PSI. Imaging in the Talbot interferometer was performed within a time frame of 2 hours from resection and were returned immediately to the clinic to proceed with the histopathological examination.

The x-ray setup consisted of an x-ray generator (Seifert ID 3000) and an unfiltered tungsten line-focus tube operated at 40 kVp with mean energy of 28 keV and a current of 25 mA. Further, a flat panel CMOS detector (Hamamatsu C9732DK) with a $12 \times 12$-cm$^2$ field-of-view and $50 \times 50$-$\mu$m$^2$ pixel size was used. A Talbot interferometer similar to the one depicted in Fig. 1 was used to acquire phase information by the phase-stepping approach. Attenuation, differential phase, and dark-field information was obtained by means of phase-retrieval.[31]

In order to simplify the evaluation task for the radiologist and to illustrate the improvement brought by high-frequency information in the phase contrast image, a dedicated fusion algorithm was developed by Philips Research and applied to the attenuation data and differential phase data after phase retrieval.[32] The algorithm is designed to achieve a noise optimal weighting of each independent Fourier component of the attenuation data and differential phase data. Its details will be published elsewhere.[33]

### 2.3. Spectral imaging

Absorption and refraction are described by the complex index of refraction; $n = 1 - \delta + i\beta$, where $\delta$ describes refraction and $\beta$ is related to absorption. With efficient scatter rejection and away from absorption edges, the real part of the refractive index and linear absorption follow approximately[34]

$$\delta \propto E^{-2}\rho \quad \text{and} \quad \mu \propto \begin{cases} E^{-3}Z^{3.2}\rho & \text{at low } E \\ \rho & \text{at high } E \end{cases}, \quad (5)$$

where $\rho$ is the mass density and $Z$ represents atomic number. Linear absorption is divided into two regions dominated by the photo-electric effect and Compton scattering at low and high energies, respectively. The crossing between the two interaction processes depends on atomic number.

When discussing energy dependence, we consider two cases. Firstly, a (ideal) setup with a monochromatic beam and design energy that follows the incident energy, i.e. $E^d = E$, by adaption of $D_n$. Hence, the influence of the setup is minimized and $A = 1$. This case is related to the intrinsic properties of phase contrast rather than specific for Talbot interferometry, and general conclusions to other phase-contrast techniques can be drawn. Secondly, we consider a setup with design energy locked at the optimal energy for phase contrast, i.e. $E^d = E^*$, which implies $A \leq 1$. This case is specific for Talbot interferometry and reflects the particular tradeoffs that are associated with a more realistic system based on the technique.

### 2.3.1. Spectral optimization

Since $\beta$ decreases monotonically with energy (except at absorption edges), the optimal incident energy in absorption contrast is affected by the tradeoff between high contrast at low energies and low noise (high transmission) at high energies.[9] $\delta$ also decreases with energy according to Eq. (5), but the dependence is weaker and we can expect the optimal incident energy to be slightly higher than for absorption contrast.

In view of Eq. (5), Eq. (2) with explicit energy dependence becomes

$$\Delta s_\Phi(E) = k(E)\Delta\delta_{cg}(E)d_c \propto E^{-1}\rho d_c \quad \text{and} \quad \Delta s_\mu(E) = \Delta\mu_{cg}(E)d_c \propto E^{-3}\rho d_c, \qquad (6)$$

where we have used $\lambda \propto E^{-1}$ and assumed dominance by the photo-electric effect. In an ideal photon-counting system, the detected number of photons are $N(E) = N_0(E)\exp[-d_b\mu_g(E)]$, where $d_b$ is the breast thickness, $\mu_g$ is the linear attenuation of breast tissue, and $N_0$ is the incident number of photons. Further, the noise is uncorrelated in an ideal system so that $S_{Q0}(E) = N(E)$. Assuming case 1 above, i.e. $E^d = E$, $\lambda^d = \lambda$, and $A = 1$, Eq. (4) yields

$$S_{Q\Phi}(E) \propto \frac{1}{N_0(E)\exp[-CE^{-3}\rho d_b]} \quad \text{and} \quad S_{Q\mu}(E) \propto \frac{1}{N_0(E)\exp[-CE^{-3}\rho d_b]}, \qquad (7)$$

where $C$ is a constant. Equation (7) shows that the phase- and absorption-contrast NPS have identical energy dependencies for this case, and there is no directional energy dependence.

If we require the dose to be equal at each energy, $N_0(E) \propto 1/D(E)$, and further assume the dose to be proportional to the reciprocal of the energy, $N_0(E) \propto 1/D(E) \propto E$ and the detectability index (Eq. (1)) becomes

$$d'^2_\Phi(E) \propto \exp(-CE^{-3}d_b) \times E^{-1}d_c^2 \quad \text{and} \quad d'^2_\mu(E) \propto \exp(-CE^{-3}d_b) \times E^{-5}d_c^2. \qquad (8)$$

A maximum of $d'^2$ with respect to energy (the optimal energy; $E^*$) can be found for instance with differentiation, i.e. by setting $\partial d'^2/\partial E = 0$, which evaluates to

$$E^*_\Phi \propto (3 \times Cd_b)^{1/3} \quad \text{and} \quad E^*_\mu \propto (3/5 \times Cd_b)^{1/3}. \qquad (9)$$

As expected, the optimal incident energy in phase contrast is slightly higher than for absorption contrast (a factor of $5^{1/3} \sim 1.7$). We further note that $E^*$ is independent of target thickness ($d_c$) and material ($\mu$), which is in agreement with previous results for absorption contrast.[35]

If we instead consider case 2 according to above, i.e. $E^d = E^*$, which is closer to a realistic system based on Talbot interferometry, the situation becomes more complicated than the general result obtained in Eq. (9). The optimal energy is additionally affected by reduced amplitude of the interference fringes away from the design energy ($E^*$) of the setup according to

$$A \propto \frac{1}{2}\left[1 + \bigwedge\left(\pi\frac{E^d}{E}n\right)\right], \qquad (10)$$

where $\bigwedge$ is the continuous triangle function, defined here as $\bigwedge(\theta) \equiv 2/\pi \times \int_0^\theta \text{sgn}[\sin(\phi)]d\phi - 1$.[22] Reduced amplitude leads to reduced fringe visibility and increased noise, which favors imaging at the design energy. There is, however, an additional energy dependence on $S_{Q\Phi}$ because $\lambda^d/\lambda$ in Eq. (4) does not cancel. This means that the increased noise towards lower energies is mitigated, whereas the increase towards higher energies is amplified. In Talbot interferometry, the phase-contrast NPS is asymmetric according to Eq. (4) so that these effects appear in only one direction, and the influence is reduced by the square root.

### 2.3.2. Energy weighting

For a given incident spectrum, detected photons can be weighted according to their information content, i.e. low-energy photons are assigned a higher weight.[12–15] For an ideal photon-counting system with several energy bins indexed by $\Omega$, the detected number of photons are $N = \sum N_\Omega = q_0 \sum \phi_\Omega$, where $q_0$ is the total number of counts and $\phi_\Omega$ accounts for the spectrum ($\sum \phi_\Omega = 1$). Equation (2) for this system becomes

$$\Delta s_\mu = |\langle I_{\mu c}\rangle - \langle I_{\mu g}\rangle| = \left|\ln\left[q_0\sum\phi_\Omega w_\Omega \exp(-\Delta\overline{\mu}_{cg\Omega} \times d_c)\right] - \ln\left[q_0\sum\phi_\Omega w_\Omega\right]\right|$$

$$= \left|\ln\left[\frac{\sum\phi_\Omega w_\Omega \exp(-\Delta\overline{\mu}_{cg\Omega} \times d_c)}{\sum\phi_\Omega w_\Omega}\right]\right| \simeq \frac{\sum\phi_\Omega w_\Omega \times \Delta\overline{\mu}_{cg\Omega}d_c}{\sum\phi_\Omega w_\Omega}, \quad \text{and}$$

$$\Delta s_\Phi = |\langle I_{\Phi c}\rangle - \langle I_{\Phi g}\rangle| \simeq \frac{\sum\phi_\Omega w_\Omega \times k_\Omega\Delta\overline{\delta}_{cg\Omega}d_c}{\sum\phi_\Omega w_\Omega}, \qquad (11)$$

where $\Delta\bar{\mu}_{\text{cg},\Omega} \equiv |\bar{\mu}_{\text{c},\Omega} - \bar{\mu}_{\text{g},\Omega}|$ and $\Delta\bar{\delta}_{\text{cg},\Omega} \equiv |\bar{\delta}_{\text{c},\Omega} - \bar{\delta}_{\text{g},\Omega}|$ are the differences in effective $\mu$ and $\delta$ over the energy bin. The weight factor $w_\Omega$ is applied before the logarithm or phase calculation. For small signal differences it would, however, be approximately equivalent to propagate the energy bins through logarithm / phase calculation before weighting.[14]

With uncorrelated noise, $S_{Q0\Omega} = N_\omega = q_0 \times \phi_\Omega$. For case 1, i.e. $E^d = E \Rightarrow \lambda^d = \lambda$ and $A = 1$, Eq. (4) becomes

$$S_{Q\mu}(\boldsymbol{f}) = \sum \left.\frac{\partial I}{\partial N_\Omega}\right|_{\overline{N}_\Omega}^2 \times S_{Q\Omega}(f) = \frac{1}{q_0} \times \frac{1}{(\sum \phi_\Omega w_\Omega)^2} \times \sum \phi_\Omega w_\Omega^2 \quad \text{and}$$

$$S_{Q\Phi}(\boldsymbol{f}) = \begin{cases} \frac{8}{\pi^2} \times \frac{1}{\Lambda^2 n^2 p_1^2} \times \frac{1}{f_x^2} \times \frac{2}{q_0} \times \frac{1}{(\sum \phi_\Omega w_\Omega)^2} \times \sum \phi_\Omega w_\Omega^2 \\ \frac{1}{q_0} \times \frac{1}{(\sum \phi_\Omega w_\Omega)^2} \times \sum \phi_\Omega w_\Omega^2 \end{cases}. \quad (12)$$

Note that the energy dependencies of $S_{Q\mu}$ and $S_{Q\Phi}$ are equal and independent of direction, similar to Eq. (7). If we combine Eqs. (11) and (12), Eq. (1) evaluates to

$$d'^2_\mu \propto \frac{(\sum \phi_\Omega w_\Omega \times \Delta\bar{\mu}_{cg\Omega} d_c)^2}{\sum \phi_\Omega w_\Omega^2} \quad \text{and} \quad d'^2_\Phi \propto \frac{(\sum \phi_\Omega w_\Omega \times k_\Omega \Delta\bar{\delta}_{cg\Omega} d_c)^2}{\sum \phi_\Omega w_\Omega^2}, \quad (13)$$

where we have assumed that the MTF is independent of energy; a fairly good approximation in our case.[11]

A global maximum of $d'^2$ with respect to weight can be found for instance by differentiation, i.e. by setting $\partial d'^2 / \partial w_n = 0$. For two energy bins (i.e. $\Omega \in lo, hi$),

$$\frac{w^*_{\Phi,lo}}{w^*_{\Phi,hi}} = \frac{\Delta\bar{\delta}_{lo} \lambda_{hi}}{\Delta\bar{\delta}_{hi} \lambda_{lo}} \Rightarrow w^*_\Phi \propto E^{-1} \quad \text{and} \quad \frac{w^*_{\mu,lo}}{w^*_{\mu,hi}} = \frac{\Delta\bar{\mu}_{lo}}{\Delta\bar{\mu}_{hi}} \Rightarrow w^*_\mu \propto E^{-3}, \quad (14)$$

where we have used Eq. (5) to evaluate energy dependence. Hence, photons used for absorption contrast should be weighted according to $E^{-3}$, which is in accordance with previous results.[13] For phase-contrast imaging in general, photons should be weighted according to $E^{-1}$. In most practical cases, however, phase-contrast efficiency is reduced away from the setup design energy, which needs to be taken into account; photons close to the design energy should be weighted higher. For Talbot interferometry, optimal weighting is affected according to the fringe amplitude in Eq. (10).

### 2.3.3. Material decomposition

X-ray attenuation in the medical imaging domain follows approximately $\mu = \mu_{PE} + \mu_C$, where $\mu_{PE}$ and $\mu_C$ are the linearly independent contributions from photo absorption and Compton scattering as given by Eq. (5).[14–17] Because of the number of interaction effects, the proportions of not more than two materials in a mixture may be determined with measurements at different x-ray energies, i.e. measurements at more than two energies are redundant. Talbot interferometry simultaneously detects absorption and phase contrast, which together with spectral imaging has the potential to add independent interaction processes for separation of more materials. In fact, a Talbot interferometer with optimal operation at several energies for inter alia this purpose has already been suggested.[36]

However, according to Eq. (5), x-ray refraction follows $\delta \propto E^{-2} \times \rho$ so that (1) all materials have the same energy dependence and (2) the material dependence of $\delta$ is limited to density and is hence correlated to Compton scattering. Therefore, (1) material decomposition is in principle not possible in phase contrast without absorption contrast and (2) phase contrast does not add information to absorption-contrast spectral imaging. The same is true for contrast-enhanced imaging since the overall energy dependencies of Compton scattering and $\delta$ are unaffected by absorption edges.

Nevertheless, phase together with absorption contrast, as obtained in e.g. Talbot interferometry, may be used as a substitute for spectral imaging. This possibility is not further pursued in the present study, but been investigated by other authors.[8]

# 3. RESULTS AND DISCUSSION

## 3.1. Non-spectral imaging

### 3.1.1. Cascaded-systems analysis

Previous results on non-spectral Talbot interferometry are summarized in this section to illustrate the framework that was introduced in Sec. 2.2.1. For more details we refer to Ref. 22.

Phase-contrast imaging did not exhibit a general signal-difference-to-noise improvement relative to absorption contrast, but the performance was found to be highly task dependent. Two of the observed effects are illustrated in Fig. 2. Firstly, the intrinsic detection of the phase differential caused correlation of the noise when integrating to phase contrast, and the NPS decreased rapidly with spatial frequency according to Eq. (4). This brown noise that shows up as streaks in Fig. 2 is less disturbing at higher spatial frequencies, and phase contrast was beneficial for small and sharp targets, e.g. tumor spicula rather than solid tumors, and for discrimination tasks rather than detection tasks. This is illustrated in Fig. 2 (a) by phase-propagation simulations of two target sizes (300 $\mu$m and 5 mm); the small target is easier to distinguish in phase contrast whereas the large target is better visualized in absorption contrast. Note that the printed pixel size varies in these two images in order to be able to display them side-by-side. Figure 2 (c) further illustrates the effect by means of the detectability benefit ratio $(d'_\Phi/d'_0)$ of phase over absorption contrast for a range of target sizes; it is evident that the benefit of phase contrast goes up for smaller radii.

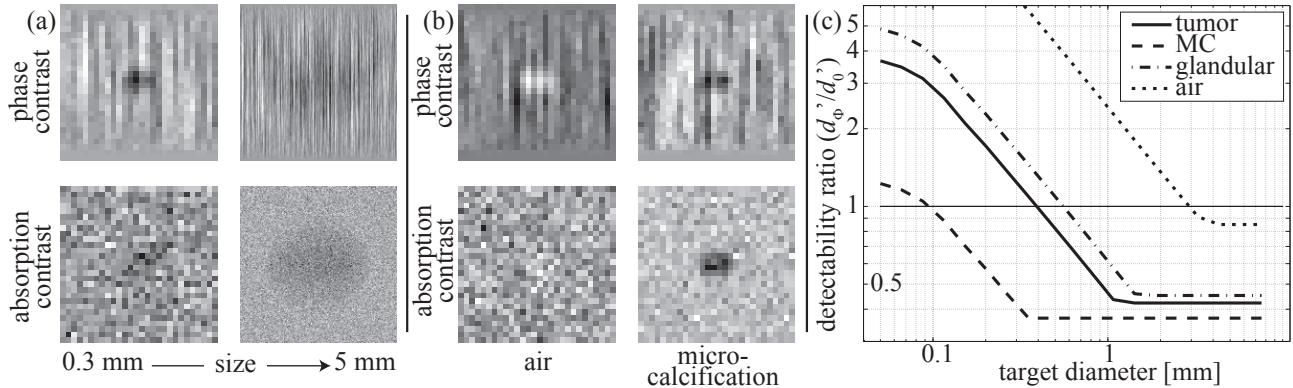

Figure 2. Results on phase-contrast size and material dependence.[22] (a) Size dependence: Simulated images of tumor structures with diameters 300 $\mu$m and 5 mm. Phase contrast (top row) and absorption contrast (bottom row). The phase-contrast NPS decreases rapidly with frequency because of integration of quantum noise, which shows up as streaks in the vertical direction. (b) Material dependence: A 300-$\mu$m-diameter air cavity (left column) and a microcalcification (right column). (c) Detectability benefit ratio of phase over absorption contrast $(d'_\Phi/d'_0)$. Printed with permission from Medical Physics.

Secondly, phase contrast favored detection of materials that differ in density compared to the background tissue, rather than materials with differences in atomic number that are efficiently probed by absorption contrast. For instance, the improvement of phase contrast in microcalcification detection was less than for tumor and glandular structures of the same size, which can be seen in Fig. 2 (c). The extreme case is a gaseous target, which is used for comparison to a 300-$\mu$m-diameter microcalcification in Fig. 2 (b).

### 3.1.2. Phase-contrast imaging of human mastectomy breast samples

Figure 3 shows images from the Talbot interferometer at PSI. It is a small section of a breast mastectomy sample from a female patient, aged 88 (case 3 in Ref. 30), with an invasive ductal breast carcinoma. The image fusion algorithm operated on the absorption image shown in the left panel of Fig. 3 and the differential phase-contrast image shown in the center panel, with the result shown in the right-hand panel. Note that the image fusion algorithm accounts for the differential nature of the data by a division in Fourier space of the Fourier amplitudes by the frequency. The improvement in the visualization of fine details and interfaces can be appreciated from a comparison between the absorption image and the fused image. We attribute this improvement to the reduced phase-contrast NPS at high frequencies as described by Eq. (4).

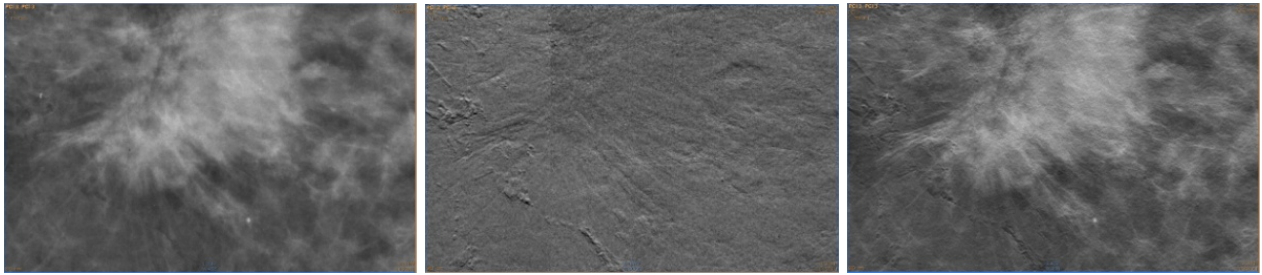

**Figure 3.** Images acquired at PSI of an invasive ductal breast carcinoma in a mastectomy sample. **Left:** Absorption-contrast image. **Center:** Differential phase-contrast image. **Right:** Image fusion of the absorption- and phase-contrast images.

## 3.2. Spectral imaging

### 3.2.1. Spectral optimization

Figure 4 shows detectability for 200-$\mu$m tumor structures and microcalcifications at 1 mGy as a function of incident photon energy. Detectability was evaluated for the two cases described in Sec. 2.3: A setup with the design energy adapted to the incident energy, i.e. $E^d = E$; Talbot interferometry with design energy locked at the optimal energy, i.e. $E^d = E^*$. Detectability for absorption contrast is plotted for comparison.

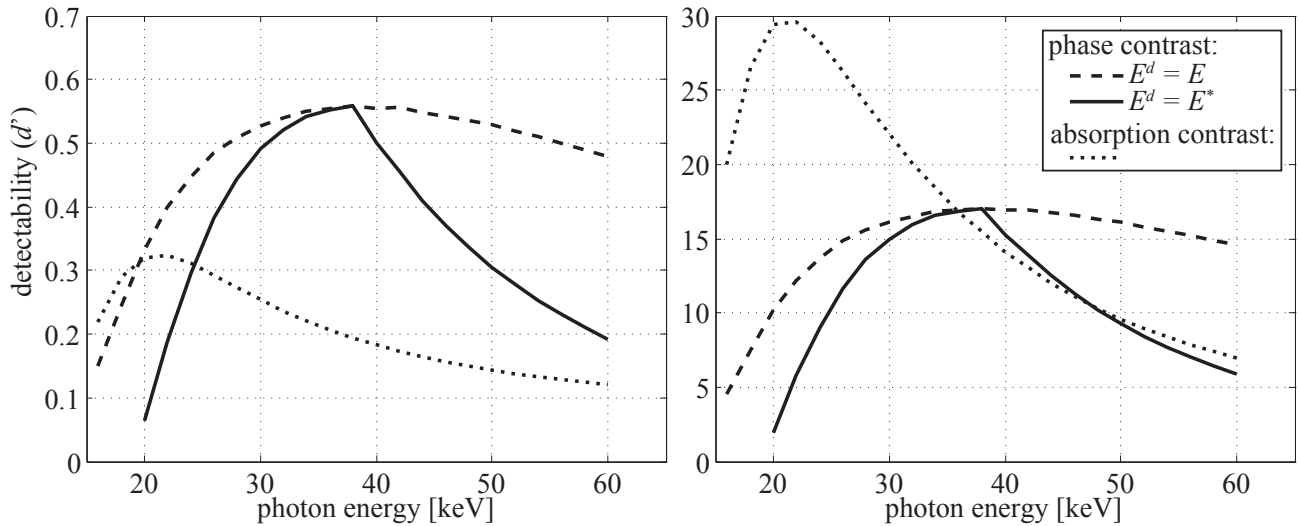

**Figure 4.** Spectral optimization: Detectability at 1 mGy as a function of energy for phase contrast without effects of the setup design energy ($E^d = E$), Talbot interferometry including the design energy ($E^d = E^*$), and absorption contrast. **Left:** Detectability of a tumor structure. **Right:** Detectability of a microcalcification.

The two typical mammography targets in Fig. 4 have similar detectability energy dependence, which is expected from Eq. (9) and is a well-known effect for absorption contrast.[35] Detectability of the microcalcification is substantially higher than for the tumor structure because of higher contrast, but there is no benefit of phase over absorption contrast. The latter is an effect of the higher atomic number and in accordance with Fig. 2 (c). The optimal energy of phase contrast is generally higher than for absorption contrast; 38 keV and 22 keV respectively according to Fig. 4. The ratio of $38/22 \sim 1.7$ matches exactly the prediction in Eq. (9).

For a setup without influence of the design energy ($E^d = E$), imaging at the optimal energy improved detectability by about 40% compared to a setup optimized for absorption contrast. If the design energy was locked, however ($E^d = E^*$), reduced amplitude modulation away from the design energy resulted in a sharper maximum, and optimization for phase contrast yielded an improvement by a factor of 3 compared to imaging at the optimum for absorption contrast.

### 3.2.2. Energy weighting

Figure 5 plots the weight that should be assigned to each photon as a function of energy to maximize detectability. The target was a 200-$\mu$m tumor structure, but almost identical behavior was found for other targets. The same cases that were considered in Fig. 4 are represented also in Fig. 5: Phase contrast with the design energy adapted to the incident energy ($E^d = E$); Talbot interferometry with design energy locked at the optimal energy ($E^d = E^*$). Optimal weighting for absorption-contrast is plotted for comparison.

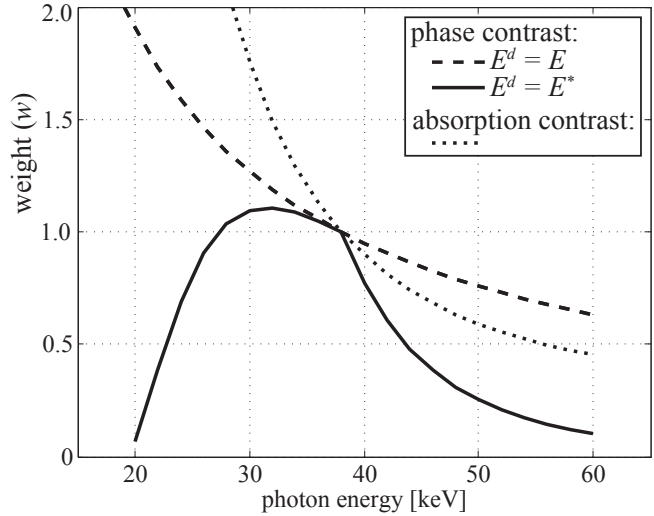

**Figure 5.** Energy weighting: Optimal weight as a function of energy for phase contrast without effects of the setup design energy ($E^d = E$), Talbot interferometry including the design energy ($E^d = E^*$), and absorption contrast.

As was discussed in Sec. 2.3.1, absorption contrast should be weighted close to $E^{-3}$. For an incident energy spectrum that was rect-distributed between 16 and 40 keV, which is a realistic energy interval for mammography, optimal weighting improved detectability by 19% compared to unweighted photon counting (intrinsic weighting $w_\mu^*(E) \propto 1$) and by 33% compared to energy integrating detectors (intrinsic weighting $w_\mu^*(E) = \propto E$). These results are in line with previous studies.[13]

Phase contrast on the other hand should be weighted according to $E^{-1}$ for a setup without influence of the design energy ($E^d = E$). Because of the weaker energy dependence, optimal weighting in phase contrast had a smaller impact than in absorption contrast; detectability was improved by 3.6% compared to photon counting and by 13% compared to energy integrating detectors with the rect-distributed incident spectrum. In Talbot interferometry with the setup design energy taken into account ($E^d = E^*$), the energy dependence was stronger, however. Optimal weighting dropped quicker towards higher energies and there was a superimposed maximum close to the design energy. Accordingly, the improvement in detectability was larger in this case; 38% compared to photon counting and 61% compared to energy integrating detectors.

# 4. CONCLUSIONS

Cascaded-systems analysis enables comprehensive evaluation of phase-contrast efficiency. The benefit compared to absorption contrast is highly dependent on task, in particular target size and material, with larger improvements for small structures, and for soft tissue rather than microcalcifications. Measurements on mastectomy samples with a conventional detector illustrated the improved detectability of fine tumor structures.

The optimal incident energy is a factor of 1.7 higher in phase contrast than in absorption contrast because the phase shift drops slower with energy than does absorption. The difference is smaller than could be expected, however, partly because Compton scattering dominates absorption contrast at higher energies. Reduced phase-contrast efficiency away from the design energy in Talbot interferometry further sharpens the optimum in incident energy, and detectability was improved by a factor of 3 compared to a setup optimized for absorption contrast.

Optimal weighting in phase contrast follows $E^{-1}$, compared to $E^{-3}$ in absorption contrast. In Talbot interferometry, the energy dependence is stronger and there is a maximum at the setup design energy. Optimal weighting improved phase-contrast detectability by 3.6–38% compared to non-spectral photon counting detectors and by 13–61% compared to energy integrating detectors.

Spectral material decomposition was not facilitated by phase contrast, but phase may be used instead of spectral information.

# ACKNOWLEDGMENTS

This research was funded in part by the Swedish agency for innovation systems (VINNOVA).

# REFERENCES

1. Lewis, R. A., "Medical phase contrast x-ray imaging: current status and future prospects," *Physics in Medicine and Biology* **49**(16), 3573–3583 (2004).
2. Zhou, S. A. and Brahme, A., "Development of phase-contrast x-ray imaging techniques and potential medical applications," *Phys Med* **24**(3), 129–48 (2008).
3. Keyrilainen, J., Bravin, A., Fernandez, M., Tenhunen, M., Virkkunen, P., and Suortti, P., "Phase-contrast x-ray imaging of breast," *Acta Radiologica* **51**(8), 866–884 (2010).
4. Clauser, J. F., "Ultrahigh resolution interferometric X-ray imaging." U.S. Patent 5,812,629 (1998).
5. David, C., Nohammer, B., Solak, H. H., and Ziegler, E., "Differential x-ray phase contrast imaging using a shearing interferometer," *Applied Physics Letters* **81**(17), 3287–3289 (2002).
6. Momose, A., Kawamoto, S., Koyama, I., Hamaishi, Y., Takai, K., and Suzuki, Y., "Demonstration of x-ray Talbot interferometry," *Japanese Journal of Applied Physics* **42**(7B), L866–L868 (2003).
7. Pfeiffer, F., Weitkamp, T., Bunk, O., and David, C., "Phase retrieval and differential phase-contrast imaging with low-brilliance x-ray sources," *Nature Physics* **2**(4), 258–261 (2006).
8. Qi, Z. H., Zambelli, J., Bevins, N., and Chen, G. H., "A novel quantitative imaging technique for material differentiation based on differential phase contrast CT," in [*Proc. SPIE, Physics of Medical Imaging*], Hsieh, J. and Samei, E., eds., **7622** (2010).
9. Motz, J. and Danos, M., "Image information content and patient exposure," *Med. Phys.* **5**(1), 8–22 (1978).
10. Engel, K. J., Geller, D., Köhler, T., Martens, G., Schusser, S., Vogtmeier, G., and Rössl, E., "Contrast-to-noise in X-ray differential phase contrast imaging," *Nucl. Instr. and Meth. A* **648, Supplement 1**(0), S202–S207 (2011).
11. Fredenberg, E., Lundqvist, M., Cederström, B., Åslund, M., and Danielsson, M., "Energy resolution of a photon-counting silicon strip detector," *Nucl. Instr. and Meth. A* **613**(1), 156–162 (2010).
12. Tapiovaara, M. and Wagner, R., "SNR and DQE analysis of broad spectrum x-ray imaging," *Phys. Med. Biol.* **30**, 519–529 (1985).
13. Cahn, R., Cederström, B., Danielsson, M., Hall, A., Lundqvist, M., and Nygren, D., "Detective quantum efficiency dependence on x-ray energy weighting in mammography," *Med. Phys.* **26**(12), 2680–3 (1999).
14. Fredenberg, E., Hemmendorff, M., Cederström, B., Åslund, M., and Danielsson, M., "Contrast-enhanced spectral mammography with a photon-counting detector," *Med. Phys.* **37**(5), 2017–2029 (2010).

15. Fredenberg, E., Åslund, M., Cederström, B., Lundqvist, M., and Danielsson, M., "Observer model optimization of a spectral mammography system," in [*Proc. SPIE, Physics of Medical Imaging*], Samei, E. and Pelc, N. J., eds., **7622** (2010).
16. Alvarez, R. and Macovski, A., "Energy-selective reconstructions in x-ray computerized tomography," *Phys. Med. Biol.* **21**, 733–744 (1976).
17. Lehmann, L. A., Alvarez, R. E., Macovski, A., Brody, W. R., Pelc, N. J., Riederer, S. J., and Hall, A. L., "Generalized image combinations in dual KVP digital radiography," *Med. Phys.* **8**(5), 659–667 (1981).
18. Johns, P., Drost, D., Yaffe, M., and Fenster, A., "Dual-energy mammography: initial experimental results," *Med. Phys.* **12**, 297–304 (1985).
19. Lewin, J., Isaacs, P., Vance, V., and Larke, F., "Dual-energy contrast-enhanced digital subtraction mammography: Feasibility," *Radiology* **229**, 261–268 (2003).
20. Baldelli, P., Bravin, A., Maggio, C. D., Gennaro, G., Sarnelli, A., Taibi, A., and Gambaccini, M., "Evaluation of the minimum iodine concentration for contrast-enhanced subtraction mammography," *Phys. Med. Biol.* **51**(17), 4233–51 (2006).
21. Bornefalk, H., Lewin, J. M., Danielsson, M., and Lundqvist, M., "Single-shot dual-energy subtraction mammography with electronic spectrum splitting: Feasibility," *Eur. J. Radiol.* **60**, 275–278 (2006).
22. Fredenberg, E., Danielsson, M., Stayman, J. W., Siewerdsen, J. H., and Åslund, M., "Cascaded-systems analysis of phase-contrast imaging," *Med. Phys.* (2012). Submitted.
23. Lundqvist, M., *Silicon strip detectors for scanned multi-slit x-ray imaging*, PhD thesis, Royal Institute of Technology (KTH), Stockholm (2003).
24. Åslund, M., Cederström, B., Lundqvist, M., and Danielsson, M., "Physical characterization of a scanning photon counting digital mammography system based on Si-strip detectors," *Med. Phys.* **34**(6), 1918–1925 (2007).
25. Sharp, P. F., Metz, C. E., Wagner, R. F., Myers, K. J., and Burgess, A. E., "ICRU Rep. 54 Medical imaging: the assessment of mage quality," *International Commission on Radiological Units and Measurements, Bethesda, MD* (1996).
26. Metz, C. E., Wagner, R. F., Doi, K., Brown, D. G., Nishikawa, R. M., and Myers, K. J., "Toward consensus on quantitative assessment of medical imaging-systems," *Med. Phys.* **22**(7), 1057–1061 (1995).
27. Siewerdsen, J. H., [*The handbook of medical image perception and techniques*], ch. 25. Optimization of 2D and 3D radiographic imaging systems, Cambridge University Press, Cambridge (2010).
28. Siewerdsen, J. H. and Jaffray, D. A., "Optimization of x-ray imaging geometry (with specific application to flat-panel cone-beam computed tomography)," *Med. Phys.* **27**(8), 1903–14 (2000).
29. Cunningham, I. A., [*Handbook of Medical Imaging*], vol. 1. Physics and Psychophysics, ch. 2. Applied Linear-Systems Theory, SPIE Press, Bellingham, USA (2000).
30. Stampanoni, M., Wang, Z. T., Thuring, T., David, C., Roessl, E., Trippel, M., Kubik-Huch, R. A., Singer, G., Hohl, M. K., and Hauser, N., "The first analysis and clinical evaluation of native breast tissue using differential phase-contrast mammography," *Investigative Radiology* **46**(12), 801–806 (2011).
31. Pfeiffer, F., Bech, M., Bunk, O., Kraft, P., Eikenberry, E. F., Bronnimann, C., Grunzweig, C., and David, C., "Hard-x-ray dark-field imaging using a grating interferometer," *Nature Materials* **7**(2), 134–137 (2008).
32. Roessl, E., Koehler, T., van Stevendaal, U., Martens, G., Hauser, N., Wang, Z., and Stampanoni, M., "Image fusion algorithm for differential phase contrast imaging," in [*Proc. SPIE, Physics of Medical Imaging*], (2012). Submitted.
33. Roessl, E. et al., "Composition algorithm for differential phase contrast projection imaging," In preparation.
34. Cederström, B., Lundqvist, M., and Ribbing, C., "Multi-prism x-ray lens," *Appl. Phys. Lett.* **81**(8), 1399–1401 (2002).
35. Fahrig, R. and Yaffe, M. J., "Optimization of spectral shape in digital mammography: dependence on anode material, breast thickness, and lesion type," *Med Phys* **21**(9), 1473–81 (1994).
36. Kottler, C. and Kaufmann, R., "Interferometer device and method." European Patent EP 2 060 909 A1 (2009).